\documentclass[twocolumn]{aastex631}

\def\code#1{\texttt{#1}}

\shorttitle{Chandra Observations of Six Peter Pan Disk Systems}
\shortauthors{Laos et al.}
\graphicspath{{./}{figures/}}
\usepackage{textalpha}

\begin{document}

\title{Chandra Observations of Six Peter Pan Disks: Diversity of X-ray-driven Internal Photoevaporation Rates Doesn't Explain Their Rare Longevity}

%%
%% Use \email to set provide email addresses. Each \email will appear on its
%% own line so you can put multiple email address in one \email call. A new
%% \correspondingauthor command is available in V6.31 to identify the
%% corresponding author of the manuscript. It is the author's responsibility
%% to make sure this name is also in the author list.
%%
%% While authors can be grouped inside the same \author and \affiliation
%% commands it is better to have a single author for each. This allows for
%% one to exploit all the new benefits and should make book-keeping easier.
%%
%% If done correctly the peer review system will be able to
%% automatically put the author and affiliation information from the manuscript
%% and save the corresponding author the trouble of entering it by hand.

%\correspondingauthor{August Muench}
%\email{greg.schwarz@aas.org, gus.muench@aas.org}

\author[0000-0001-8407-2105]{Stefan Laos}
\affiliation{Homer L. Dodge Department of Physics and Astronomy, University of Oklahoma
\\
440 West Brooks Street, Norman, OK 73019, USA}

\author[0000-0001-9209-1808]{John P. Wisniewski}
\affiliation{Homer L. Dodge Department of Physics and Astronomy, University of Oklahoma
\\
440 West Brooks Street, Norman, OK 73019, USA}

\author[0000-0002-2387-5489]{Marc J. Kuchner}
\affiliation{NASA Goddard Space Flight Center, Exoplanets and Stellar Astrophysics Laboratory, Code 667, Greenbelt, MD 20771, USA}

\author{Steven M. Silverberg}
\affiliation{MIT Kavli Institute for Astrophysics and Space Research, 77 Massachusetts Avenue, Cambridge, MA 02139, USA}

\author[0000-0003-4243-2840]{Hans Moritz Günther}
\affiliation{MIT Kavli Institute for Astrophysics and Space Research, 77 Massachusetts Avenue, Cambridge, MA 02139, USA}

\author[0000-0002-7939-377X]{David A. Principe}
\affiliation{MIT Kavli Institute for Astrophysics and Space Research, 77 Massachusetts Avenue, Cambridge, MA 02139, USA}

\author{Brett Bonine}
\affiliation{Homer L. Dodge Department of Physics and Astronomy, University of Oklahoma
\\
440 West Brooks Street, Norman, OK 73019, USA}

\author[0000-0002-5365-1267]{Marina Kounkel}
\affiliation{Department of Physics and Astronomy, Vanderbilt University, VU Station 1807, Nashville, TN 37235, USA}

\author{The Disk Detective Collaboration}



\begin{abstract}

We present \textit{Chandra} X-ray observations of 6 previously-identified Peter Pan objects, rare $\sim$40 Myr systems with evidence of primordial disk retention. We observe X-ray luminosities (0.8-3.0 keV) ranging from log $L_{x}$ $\sim$ 27.7--29.1. We find that our Peter Pan sample exhibits X-ray properties similar to that of weak-lined T-Tauri stars and do not exhibit evidence of stellar accretion induced X-ray suppression. Our observed Peter Pan X-ray luminosities are consistent with that measured for field dM stars of similar spectral type and age, implying their long primordial disk lifetimes are likely not a consequence of unusually faint X-ray host stars. Our derived X-ray photoevaporative mass loss rates predict our systems have passed the point of rapid gas dispersal and call into question the impact of this internal mechanism for primordial disk dispersal around dM stars. Our qualitative assessment of the surrounding Peter Pan environments also does not predict unusually low levels of external photoevaporation relative to other respective moving group members. Overall, our results suggest Peter Pan disks may be a consequence of the low FUV flux incident on the disk in low-mass DM stars given their relatively lower levels of accretion over the course of their pre-main-sequence evolution.

%Our qualitative assessment of the surrounding environment of our Peter Pan sample finds evidence of current relative isolation to other moving group massive members, potentially hinting at an overall low level of external photoevaporation experienced throughout their pre-main-sequence evolution. If external photoevaporation is indeed critical for dM stars, as our investigation of X-ray driven internal photoevaporation could suggest, this might explain why Peter Pan disks have persisted to such long lifetimes and their overall rarity compared to other (disk-less) dM stars.

\end{abstract}

\section{Introduction} \label{sec:intro}

Large, sensitive IR surveys conducted by NASA’s \textit{Spitzer} and \textit{WISE} satellites have demonstrated both that primordial protoplanetary disks of dust and gas clear on rapid ($\lesssim$10 Myr) timescales, and that this dissipation timescale depends on the mass of the host protostar (\citealt{2006ApJ...651L..49C}; \citealt{2010ApJ...724..835W}; \citealt{2016MNRAS.461..794P}). Although dispersion exists between different environments, the FEPS and c2d legacy \textit{Spitzer} surveys have found that $\sim$10 Myr is an upper limit for the lifetimes of primordial disks around solar-type stars (\citealt{2006ApJ...651L..49C}; \citealt{2010ApJ...724..835W}; \citealt{2011ARA&A..49...67W}). For lower-mass K-type stars, the frequency of primordial disks is $\sim$9\% at 10 Myr and $\sim$4\% by 16 Myr, indicating a disk e-folding timescale of $\sim$4-5 Myr \citep{2016MNRAS.461..794P}. These disk lifetimes not only set the timescale for the star formation process but also necessarily constrain the timescale for planet formation in these systems. Understanding the detailed evolution of primordial disks around the most abundant low-mass stars (e.g., M dwarfs) is of strong current interest, due in part to the expected windfall of M dwarf planet science from NASA’s TESS mission. TESS has already confirmed 18 planets having radii less than 2 R$_{\oplus}$ surrounding low-mass M dwarfs (e.g.~\citealt{2019ApJ...883L..16C}, \citealt{2021A&A...653A..41D}). %TESS is expected to detect $\sim$377 planets having radii less than 2 R$_{\oplus}$ surrounding low-mass M dwarfs, including 70 such terrestrial planets that reside in the habitable zone surrounding these M dwarfs \citep{2018ApJS..239....2B}.

Once depleted of gas, primordial disks evolve into second-generation debris disks. They are composed primarily of dust, created by continuous collisional destruction of cometary and asteroidal bodies that are remnants of the star and planet formation process. This destruction is thought to be at least partially stirred by the presence of planetary bodies in the system \citep{2008ARA&A..46..339W}. These systems have depleted most of their gas and no longer exhibit signatures of accretion. The frequency of debris can be as high as 20\% for young main sequence A-type stars, and such disks can persist for hundreds of Myrs \citep{2005ApJ...620.1010R}. By contrast, \citealt{2016IAUS..314..159B} found zero M dwarf debris disks in their ALLWISE study of moving groups older than 40 Myr, consistent with the small number of candidate M dwarf debris disk systems identified by \citealt{2014ApJ...794..146T} (of which some were false-positive detections, \citealt{2018ApJ...868...43S}). The observed low frequency of old ($\geq$40 Myr) M dwarf debris disks may indicate that such systems are ultimately cleared by stellar wind \citep{2008ARA&A..46..339W} or stellar activity (e.g. AU Mic; \citealt{2020ApJ...889L..21G}) over shorter timescales.

Over the past few years, a new population of low-mass, M-type stars, with ages $\sim$45 Myr reliably secured via moving group membership, have been found to exhibit substantial (L$_{IR}$/L$_{*}$ $\sim$0.1) mid-IR excesses, indicative of the presence of warm dust disks (\citealt{2016ApJ...830L..28S}; \citealt{2016ApJ...832...50B}; \citealt{2018MNRAS.476.3290M}; \citealt{2020MNRAS.494...62L}; \citealt{2020ApJ...890..106S}). While such systems were initially interpreted to be (rare) examples of the oldest dM-type debris disk systems \citep{2016ApJ...830L..28S}, their large IR excesses suggest there should be gas in the system, and is less consistent with the interpretation of a debris disk origin. Indeed, growing evidence shows some of these systems exhibit clear accretion tracers, with strong (10-125 \AA), broad (200-350 km/s) Hα emission (\citealt{2016ApJ...832...50B}; \citealt{2018MNRAS.476.3290M}; \citealt{2020ApJ...890..106S}) and variable Paschen $\beta$ and Brackett $\gamma$ emission \citep{2020ApJ...890..106S} in the canonical case of WISEA J080822.18–644357.3. With evidence of ongoing gas accretion despite their significantly advanced ages, this set of objects have been coined as ``Peter Pan" objects, given their refusal to ``grow up" \citep{2020ApJ...890..106S}. These systems may also be the pre-cursors to notable systems like TRAPPIST-1 \citep{2017Natur.542..456G}, in which a long-surviving disk around a low-mass dM star circularizes numerous planetary orbits into a stable configuration.

Overall, these systems currently challenge the standard paradigm for primordial disk (and planet formation) lifetimes. \citealt{2020ApJ...890..106S} argue the simplest explanation for these systems is that they are primordial disks that are still in the process of dissipating, implying lower than average disk gas-mass loss rates. A key driver of primordial-disk dissipation is photoevaporation of the disk’s gas content from high energy photons radiated by the host star (\citealt{1994ApJ...428..654H}; \citealt{2001MNRAS.328..485C}; \citealt{2004MNRAS.354...71A}; \citealt{2004ApJ...607..890F}; \citealt{2009ApJ...699.1639E}; \citealt{2009ApJ...690.1539G}; \citealt{2010MNRAS.401.1415O}; \citealt{2011ARA&A..49...67W}). Studies of the low-mass population of the TW Hya association have shown an anti-correlation between X-ray luminosity and disk fraction as a function of spectral type; earlier M dwarfs have higher X-ray luminosities and a lower disk fraction than mid-late M dwarfs \citep{2016AJ....152....3K}. The fundamental question remains: are the disks around Peter Pan stars primordial and if so, why have they persisted around these dwarf M stars and not others? Do Peter Pan objects represent a rare class of X-ray faint stars? 

We investigate these open questions with our new X-ray observations, which gauge the soft (0.8-3.0 keV) X-ray photons radiating from the host star. In this paper, we characterize the X-ray properties of identified Peter Pan objects and estimate their expected mass loss rates from X-ray driven photoevaporation for the first time. In Section~\ref{sec:obs_data}, we describe the nature of our \textit{Chandra} observations as well as our utilized data reduction and spectral extraction techniques. In Section~\ref{sec:results}, we report our derived Peter Pan X-ray luminosities, comparing to both active, young stars and also other dwarf M-type stars in nearby moving groups. We also compare our findings to the X-ray luminosities used in the recent modeling of \citealt{2022MNRAS.509...44W}. In Section~\ref{sec:discussion}, we interpret these observations in the context of recent modeling and proposed disk dispersal processes. Finally, in Section~\ref{sec:summary_conclusion}, we conclude with a brief summary of our findings. 
\section{Observations and Data Reduction}\label{sec:obs_data}

X-ray observations (Table~\ref{tab:xrayobs}) of our Peter Pan sample were obtained with the Advanced CCD Imaging Spectrometer (ACIS) onboard the \textit{Chandra} X-ray Observatory. All data were acquired on ACIS-S3 chip in VFAINT mode. 

To extract our X-ray spectra, we analyzed the pipeline-processed data files provided by the \textit{Chandra} X-ray Center using standard science threads with CIAO version 4.13 and calibration data from CALDB version 4.9.5 \citep{2006SPIE.6270E..1VF}. Spectra were extracted using circular apertures centered on the stellar X-ray sources, with diameters of 4\arcsec~in the case of J0501 and J0808. In the case of close binaries J0949 (a$\sim$1.5\arcsec) and J0446 (a$\sim$2.3\arcsec), we closely inspect the binned level 2 event files by eye, adopting optimal aperture diameters of $\sim$1.4\arcsec and $\sim$2.6\arcsec, respectively. These sizes minimized companion contamination while also maximizing the enclosed energy fraction given the \textit{Chandra} on-axis point spread function (PSF). For each observation, we sample the background with a circular aperture in a nearby region on the same CCD to each source. 

To derive X-ray luminosities for our Peter Pan sample, we analyze our observed X-ray spectra (Section~\ref{sec:obs_data}) using version 4.13 of CIAO/SHERPA \citep{2007ASPC..376..543D} and version 4.9.5 of CALDB \citep{2006SPIE.6270E..1VF}. We note our X-ray count data peaks around 1 keV for all objects, expected from their young, pre-main-sequence status. For each individual observation, we group the data to 5 counts per bin and consider counts in the 0.8-3 keV range for fitting. For our spectral model, we adopt the XSPEC optically thin plasma model \code{vapec}, considering only a single temperature component to avoid over-fitting our low-count data. We also add a multiplicative photoelectric absorption component \code{wabs} to our spectral model. We find negligible intervening absorption, expected given the close proximity of our sources ($\lesssim$100pc), and choose to freeze its value to zero. 

Similar to the analysis done in \citet{2016AJ....152....3K}, we fix our plasma abundances to the typical values determined for T Tauri stars in Taurus (\citealt{2013ApJ...765....3S} and references therein). Relative to solar, these abundance values are H = 1.0, He = 1.0, C = 0.45, N = 0.79, O = 0.43, Ne = 0.83, Mg = 0.26, Al = 0.50, Si = 0.31, S = 0.42, Ar = 0.55, Ca = 0.195, Fe = 0.195, and Ni = 0.195. In the case of J0446A/B and J0808, we perform simultaneous fits over the numerous observations of each object. To gauge the quality of our fits, we consider the Cash statistic \citep{1979ApJ...228..939C} as implemented in SHERPA rather than a chi-squared metric, which is not appropriate for the Poisson-distributed low-count regime of our data. The resulting best fit parameters derived from this analysis are given in Table~\ref{tab:specparams}. From these best fits, we derive X-ray fluxes in the 0.3-8 keV energy band and use \textit{Gaia} DR3 distances \citep{2021A&A...649A...1G} to derive our Peter Pan X-ray luminosities ($L_{x}$), also reported in Table~\ref{tab:specparams}. Our best fits are shown in Appendix Section~\ref{sec:spec_fitting}. Given the low count rates and correspondingly large relative Poisson uncertainty, we note we cannot exclude variability of a factor of two between our observations, but would have detected order-of-magnitude flares had they occurred during our observations. 

To compute overall fractional luminosities, we also compute the bolometric luminosity of our Peter Pan sources. Following the methodology of \citet{2016AJ....152....3K}, we use known spectral types, \textit{Gaia} DR3 distances, and J band data based on the (spectral-type-dependent) J band bolometric corrections determined by \citet{2013ApJS..208....9P} assuming no reddening. For spectral types later than M5, we linearly extrapolate from the bolometric J correction relation. In the cases of J0446 and J0949, we assume the flux contribution from each stellar component is equal. These values are reported in Table~\ref{tab:specparams}.

\begin{table*}[h]
\centering
\caption{Peter Pan X-ray Observations}
\label{tab:xrayobs}
\begin{tabular}{cccc}
\hline \hline
Object & Obs. Date (UTC) & Obs. ID & Exposure (ks) \\
\hline
WISEA J094900.65–713803.1 A/B & 2019-12-13 15:34:31 & 22305 & 17 \\
2MASS J05010082–4337102 & 2019-12-22 5:15:02 & 22306 & 19 \\
WISEA J044634.16–262756.1 A/B & 2020-05-20 4:48:40 & 22304 & 22 \\
 & 2020-05-21 8:57:52 & 23255 & 10 \\
 & 2020-06-02 4:27:50 & 23256 & 23 \\
WISEA J080822.18–644357.3 & 2021-07-06 14:29:42 & 22303 & 36 \\
 & 2021-07-07 6:24:58 & 24756 & 35 \\
\hline
\end{tabular}
\end{table*}

\begin{table*}[h]
\caption{X-ray Spectral Analysis: Best-fit Parameters}
\label{tab:specparams}
\centering
\begin{tabular}{ccccc|cccc}
\hline \hline
Source & F$_{x}$ & kT & norm & Reduced Cstat\tablenotemark{a} & L$_{x}$ & L$_{bol}$ & Spectral Type\tablenotemark{b} & Age\tablenotemark{b} \\ \hline
 & erg/s/cm$^{2}$ & keV &  &  & erg/s & erg/s &  & Myr \\
 \hline
J0949A & 1.48$\pm$0.4E-13 & 0.28$\pm$0.02 & 3.0E-4$\pm$7.5E-5 & 1.69 & 1.1E+29 & 2.18E+32 & M4 & 45 \\
J0949B & 6.55$\pm$0.9E-14 & 0.73$\pm$0.13 & 8.4E-5$\pm$1.5E-5 & 1.39 & 4.8E+28 & 1.92E+32 & M5 & 45 \\
J0501 & 2.14$\pm$0.6E-14 & 0.63$\pm$0.28 & 2.0E-5$\pm$9.2E-6 & 1.49 & 5.9E+27 & 2.60E+31 & M4.5 & 42 \\
J0446A & 4.39$\pm$0.4E-14 & 0.95$\pm$0.09 & 3.8E-5$\pm$3.4E-6 & 1.05 & 3.6E+28 & 5.24E+31 & M6 & 42 \\
J0446B & 3.03$\pm$0.9E-14 & 0.50$\pm$0.10 & 2.7E-5$\pm$6.3E-6 & 1.65 & 2.5E+28 & 5.19E+31 & M6 & 42 \\
J0808 & 1.21$\pm$0.4E-14 & 0.46$\pm$0.12 & 1.1E-5$\pm$3.7E-6 & 0.77 & 1.5E+28 & 3.08E+31 & M5 & 45 \\
\hline
\end{tabular}
\tablenotetext{a}{The Cstat procedure is the XSPEC\footnote{\url{http://heasarc.gsfc.nasa.gov/xanadu/xspec.}} implementation of the Cash statistic \citep{1979ApJ...228..939C}, which also incorporates fitting a model to the background. The reduced cstat values correspond to the observed statistic divided by the degrees of freedom, operating as an approximate goodness of fit metric with good fits to the data approaching a nominal value of 1.}
\tablenotetext{b}{Spectral types are derived from the analysis of \citealt{2020ApJ...890..106S} and references therein. Ages are estimated from the isochronal fitting of \citealt{2015MNRAS.454..593B}.}
\end{table*}
\section{Results} \label{sec:results}

With our derived luminosities, we explore the X-ray properties of our 6 Peter Pan targets for the first time. In Section 3.1, we consider our X-ray luminosities in the context of stellar accretion. In Section 3.2, we compare our derived values to sensitive X-ray observations of our M star analog control sample (Appendix Section~\ref{sec:xraycomp_sample}). We also compare our luminosities to that utilized in the prescriptions of recent modeling in Section 3.3.

\subsection{\texorpdfstring{Accretion vs. L$_{x}$}{Accretion vs. Lx}}\label{subsec_accretion_results}

We begin by comparing our sample with the large sample of classical and weak-lined T-Tauri stars (CTTS, WTTS) in Taurus from \citealt{2007A&A...468..425T}. The sensitive XMM-Newton observations of this young (average age of $\sim$2.4 Myr), active sample allow us to further gauge the impact of active stellar accretion on $L_{x}$.  

\begin{figure}
    \includegraphics[width=\linewidth]{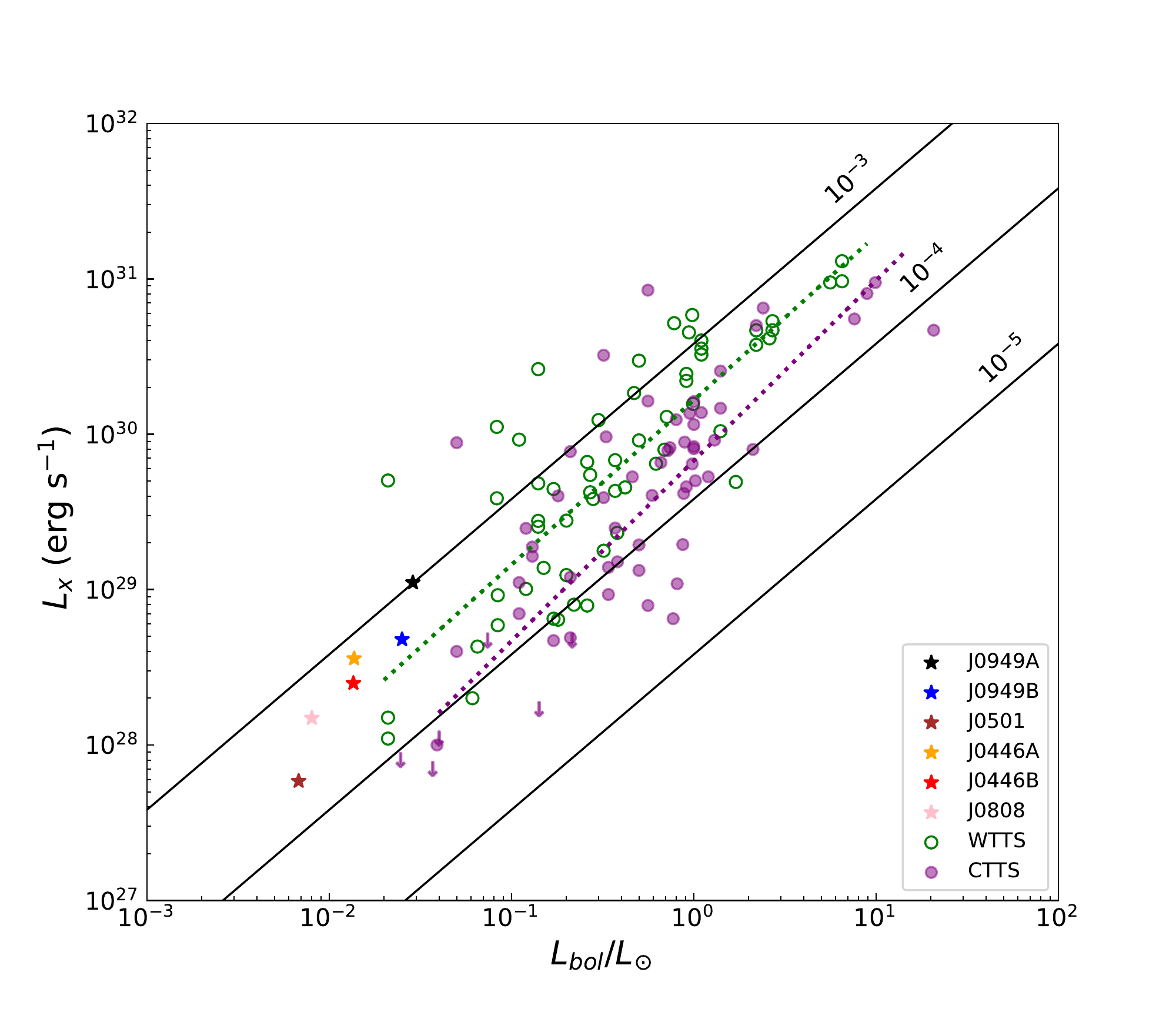}
    \caption{Comparison of observed X-ray luminosities between our Peter Pan sample (starred points) and the T Tauri stars from \citealt{2007A&A...468..425T}. Upper limits are indicated with arrows. Diagonal black lines represent constant X-ray to bolometric luminosity fractions (L$_{x}$/L$_{bol}$) of 10$^{-3}$, 10$^{-4}$, 10$^{-5}$. Linear regressions for the WTTS (green) and CTTS (purple) subsamples are reproduced from \citealt{2007A&A...468..425T}, shown as dotted lines. We find the Peter Pan sources more closely resemble the L$_{x}$/L$_{bol}$ behavior exhibited by the WTTS (green) relative to the X-ray fainter CTTS (purple).}
    \label{fig:TTS_look}
\end{figure}

Probing even lower bolometric luminosities, our observations find a large diversity in $L_{x}$ across the Peter Pan sample (Figure~\ref{fig:TTS_look}), spanning a range of $\sim$1.5 dex. Their overall appreciable X-ray luminosities do not show evidence for the significant ($\sim$2x) X-ray suppression seen in the (actively accreting) CTTS subsample \citep{2007A&A...468..425T}. We instead find the $L_{x}$/$L_{bol}$ values of the Peter Pan sources more closely follow the trend observed in the WTTS, consistent with the similarly weak level of Peter Pan accretion ($\sim$10$^{-11}$-10$^{-9}$ M\textsubscript{\(\odot\)}~yr$^{-1}$) derived from recent H-alpha observations \citealt{2020ApJ...890..106S}.  

We also compare the H-alpha emission equivalent widths from the measurements of \citealt{2020ApJ...890..106S} with our observed X-ray luminosities. We do not find evidence of a strong correlation (computing a Pearson correlation coefficient value of -0.45) within the limitations of our small sample. Interestingly, we note the strongest accretor (J0949A) is also our brightest X-ray object, counter to the expectation from stellar accretion induced X-ray suppression. Collectively, these results do not find evidence of significant X-ray suppression in the Peter Pan sample as seen in the active CTTS, likely due to their relatively low level of accretion.

%This will up come earlier
%\citep{2007A&A...468..425T} found a $\sim$2x decrease in the overall X-ray luminosity of strong (classical) accretors relative to that of weak accretors in their large sample of young T-Tauri stars.

% \begin{figure}
%     \includegraphics[width=\linewidth]{figures/halphaEW_Lx.pdf}
%     \caption{Comparison of Peter Pan H-alpha equivalent widths and X-ray luminosity. Given the lack of an observed correlation, we do not find evidence of accretion-induced X-ray suppression for our sample. 1$\sigma$ errors are shown in light red. Reported H-alpha errors (\citealt{2016ApJ...832...50B}, \citealt{2020ApJ...890..106S}) are smaller than the points ($\lesssim$0.3\AA) except in the case of J0808 \citep{2018MNRAS.476.3290M}.}
%     \label{fig:halpha_xray}
% \end{figure}

\subsection{M Star Control Sample Comparison}\label{subsec_mstar_results}

Our comparison with the T Tauri stars from \citep{2007A&A...468..425T} in Section~\ref{subsec_accretion_results} mainly considers relatively young stars ($\sim$1-10 Myr). Given the suspected advanced ages of the Peter Pan sources ($\sim$45 Myr, \citealt{2020ApJ...890..106S}), we construct a new comparison sample, closely matching suspected spectral types and ages with similarly sensitive X-ray observations. We describe our control sample criteria in Appendix Section~\ref{sec:xraycomp_sample}.

In Figure~\ref{fig:Mstarcomp}, we compare both X-ray (left panel) and fractional X-ray (right panel) luminosities between our Peter Pan sources and this control sample. Overall, we find the values observed are consistent with one another. We therefore do not find evidence that these known Peter Pan sources are over- or under-luminous in X-rays relative to similarly evolved M stars. This finding may argue against the previously thought rarity of disk longevity out to Peter Pan ages for mid-M types. We explore this possibility further in Section~\ref{sec:discussion}. 

\begin{figure*}[!ht]
    \includegraphics[width=\linewidth]{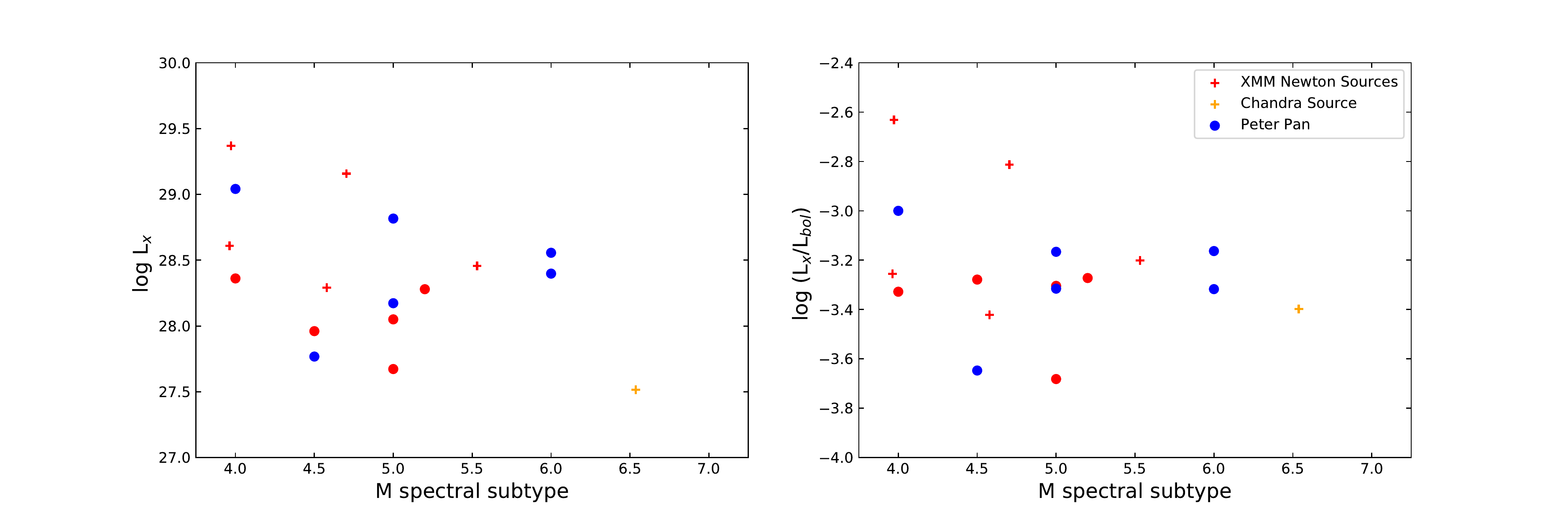}
    \caption{Observed X-ray (left) and fractional X-ray (right) luminosities as a function of spectral type between the Peter Pan sample and M star control sample (i.e., similar ages and spectral types of our Peter Pan sample). Circular points represent spectroscopically verified spectral types while crosses represent spectral types estimated via colors. Overall, we find our derived Peter Pan luminosities to be consistent with that observed for other moving group M stars.}
    \label{fig:Mstarcomp}
\end{figure*}

\subsection{Predicted vs. Observed X-ray Luminosity Evolution in dM Stars}\label{subsec_modelcomp_results}

Our X-ray observations offer a set of measurements to benchmark the characteristic X-ray luminosity for Peter Pan disk systems and assess previously used predictions. Recent modeling has investigated the survival of primordial circumstellar disks out to the advanced ages of 45 Myr (\citealt{2022MNRAS.509...44W}), assuming disk winds from strong stellar X-rays constitute the primary internal source of disk dispersal. Detailed simulations of this mass loss procedure (\citealt{2012MNRAS.422.1880O}; \citealt{2019MNRAS.487..691P}) have found a power-law dependence on the X-ray luminosity of the host star (which in turn is dependent on stellar mass). 

To estimate the expected Peter Pan X-ray luminosity, \citealt{2022MNRAS.509...44W} consider the characteristic (quiescent) X-ray luminosity-mass relation derived by \citealt{2012A&A...548A..85F} for young stars observed by the \textit{Chandra} Orion Ultra Deep Project (COUP, \citealt{2005ApJS..160..319G}), which is nearly complete ($\sim$95\%, \citealt{2005ApJS..160..401P}) down to the stellar mass limit. These X-ray luminosities steadily decrease throughout pre-main-sequence evolution, estimated to follow a time dependence of L$_{x} \propto t^{-2/5}$ (\citealt{2016MNRAS.457.3836G}; \citealt{2021A&A...649A..96J}). Together, these results give the functional form given in Equation~\ref{Lx_WZ}, where a and b are constants 30.0 and 1.87 respectively (details noted in Section 2.2 of \citealt{2022MNRAS.509...44W}). 

\begin{equation}\label{Lx_WZ}
L_x(t) = 10^{a + (b * \text{log} [M_{*} / M\textsubscript{\(\odot\)}])} \left (\frac{t}{1 \text{Myr}} \right)^{-2/5}
\end{equation}

\begin{figure}
    \includegraphics[width=\linewidth]{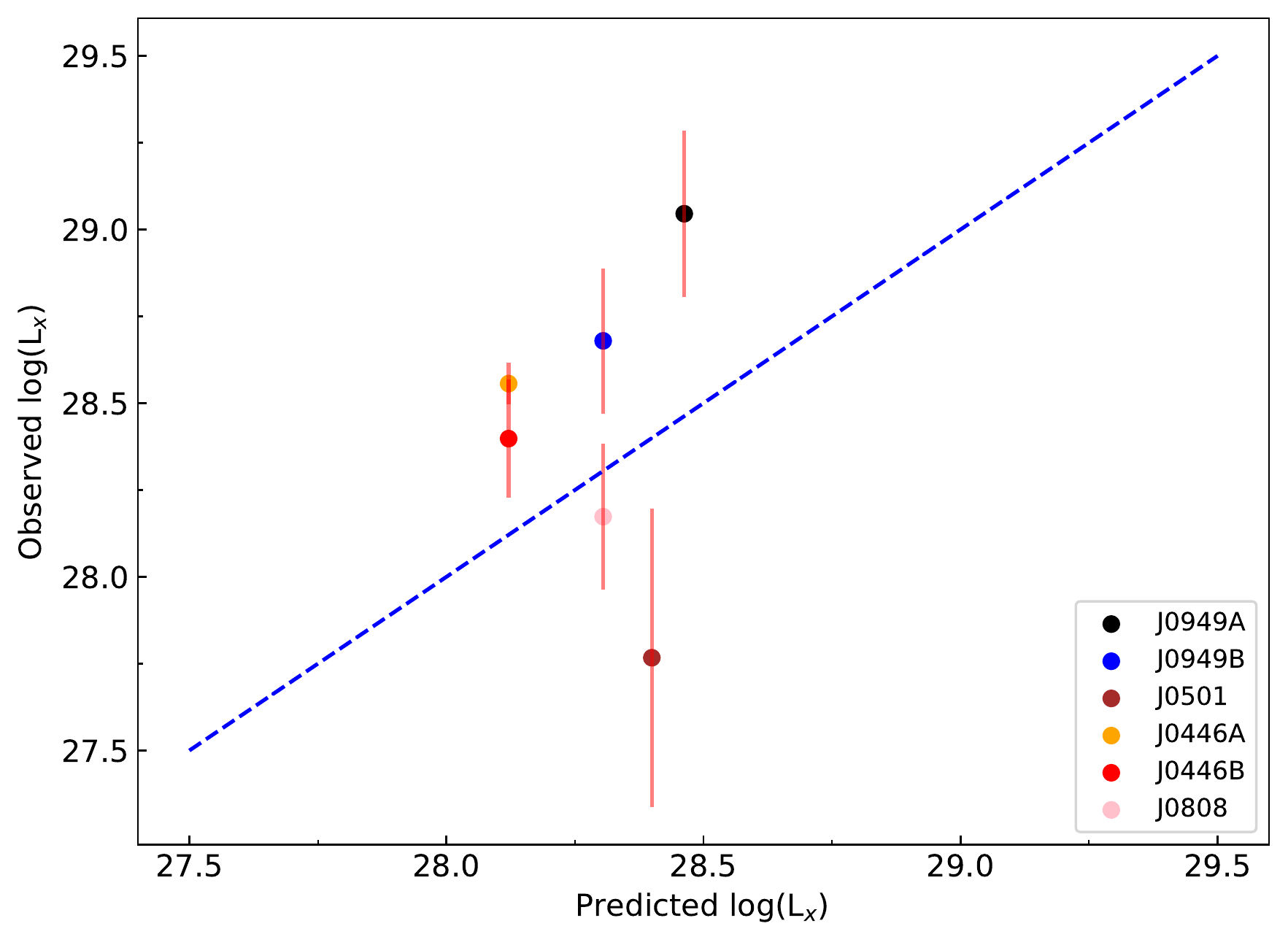}
    \caption{Comparison of observed L$_{x}$ against the predicted L$_{x}$ values used in the recent modeling conducted by \citealt{2022MNRAS.509...44W} (Section~\ref{subsec_modelcomp_results}). The dotted line shown represents the 1-1 relation. 1$\sigma$ error bars are shown in light red.}
    \label{fig:WZ_comp}
\end{figure}

In Figure~\ref{fig:WZ_comp}, we compare our observed Peter Pan X-ray luminosities with the predictions from Equation 1, assuming main sequence masses and the nominal ages estimated for the Peter Pan systems in \citealt{2020ApJ...890..106S}. Overall, we find this prescription roughly estimates the locus of our measurements but in some cases differs by up to $\sim$0.5 dex in the cases of our brightest (J0949A) and faintest (J0501) Peter Pan source. We note however that a more robust comparison is difficult, given the large uncertainties on our derived x-ray luminosities and the small size of our sample. This is further complicated by the large intrinsic scatter in observed stellar x-ray luminosities, varying by factor of $\sim$2-3 as a consequence of variations in surface magnetic activity, magnetic cycles, or flares (\citealt{2005ApJS..160..401P}, \citealt{2021A&A...649A..96J}). Although we find no evidence of flares in our x-ray light curve data, it remains unclear the extent to which Peter Pan x-ray luminosities are being over- or under- predicted in recent modeling. We discuss the implications of this result and investigate their expected mass loss rates further in Section~\ref{subsec_internal_disc}. 

% Detailed simulations of this mass loss procedure (Owen and Picogna) are in general agreement at the low suspected masses of the Peter Pan disk systems ($\sim\leq$0.3M\textsubscript{\(\odot\)}), finding a power-law dependence on the X-ray luminosity of the host star with an index of 1.14. \citealt{2022MNRAS.509...44W} consider the characteristic (quiescent) X-ray luminosity dependence on stellar mass from the young stars in Orion observed by the \textit{Chandra} Orion Ultra Deep Project (COUP, \citealt{2005ApJS..160..319G}), which is nearly complete ($\sim$95\%, \citealt{2005ApJS..160..401P}) down to stellar mass limit.  then time-evolved to match the suspected advanced ages of the Peter Pan systems. The overall trend of decreasing X-ray luminosity is well-documented over large samples of     

% mass loss rate computation will need to be included here? along with expected disk lifetime
\section{Discussion} \label{sec:discussion}

Primordial disks are subject to mass loss through a variety of mechanisms, both internal and external to the system (\citealt{2011ARA&A..49..195A}; \citealt{2011ARA&A..49...67W}). In the inner disk, these mechanisms include both the accretion of disk material onto the stellar surface \citep{2016ARA&A..54..135H} as well as disk outflows driven by the high energy flux from the central star, a process known as internal photoevaporation (\citealt{1994ApJ...428..654H}; \citealt{2001MNRAS.328..485C}; \citealt{2004MNRAS.354...71A}; \citealt{2004ApJ...607..890F}; \citealt{2009ApJ...699.1639E}; \citealt{2009ApJ...690.1539G}; \citealt{2010MNRAS.401.1415O}; \citealt{2011ARA&A..49...67W}). Photoevaporation is also capable of externally triggering mass loss in the outer disk as a consequence of the FUV radiation from nearby massive stars (\citealt{1999ApJ...515..669S}; \citealt{2001MNRAS.325..449S}; \citealt{2004ApJ...611..360A}; \citealt{2019MNRAS.490.5678C}; \citealt{2019MNRAS.485.1489W}). 

The relative importance of these internal and external channels driving disk dissipation are suspected to depend on stellar mass. Higher mass ($\gtrsim$1M\textsubscript{\(\odot\)}) stars have similarly higher levels of accretion (\citealt{2006ApJ...639L..83A}; \citealt{2014A&A...561A...2A}) and suspected internally driven photoevaporation (\citealt{2012MNRAS.422.1880O}). These mechanisms likely dominate in this case, given primordial disks around solar-type stars dissipate by $\sim$10 Myr regardless of their diverse surrounding environments (\citealt{2006ApJ...651L..49C}; \citealt{2010ApJ...724..835W}; \citealt{2016MNRAS.461..794P}). In the lower mass ($\lesssim$0.3M\textsubscript{\(\odot\)}) regime of the Peter Pan stars, the impact of external photoevaporation becomes more comparable even in low external radiation fields \citep{2018MNRAS.481..452H} given their lower internal accretion. Overall, the relative impact of internal and external mechanisms remain debated for the lowest-mass stars, with these respective class of mechanisms likely dominating disk dispersal at different points of pre-main-sequence evolution (\citealt{2022MNRAS.509...44W}). We discuss the implications of our X-ray results in regards to the described internal and external mechanisms in Section~\ref{subsec_internal_disc} and \ref{subsec_external_disc}, respectively. 

% Our X-ray observations allow for a more comprehensive assessment of the Peter Pan disk environment, complementing previously obtained near-IR/optical \citet{2020ApJ...890..106S} and radio \citep{2019ApJ...872...92F} measurements. In particular, these observations gauge the hard (1 keV) photons radiating from the host star, suspected to be a primary source of internal disk photoevaporation. We discuss the implications of our results in regards to this mechanism in Section~\ref{subsec_internal_disc}. 

% Alternatively, radiation from nearby massive stars can serve as an additional, external mechanism for disk dispersal. Given that recent modeling (\citealt{2020MNRAS.496L.111C}, \citealt{2022MNRAS.509...44W}) has required unusually low external photoevaporation to reproduce Peter Pan like-disks, we gauge the surrounding environment of our Peter Pan sample to explore this possibility in Section~\ref{subsec_external_disc}.

\subsection{The Impact of Internal Disk Dispersal Mechanisms on Peter Pan Disk Longevity}\label{subsec_internal_disc}

%The current evidence of primordial gas disk systems surviving to the advanced ages of the Peter Pan systems presents new challenges for disk evolution models, which almost ubiquitously predict gas dispersal within $\sim$10 Myr. Proposed pathways for internal disk dispersal have mainly focused on disk outflows driven by the high energy flux from the central star, a process known as photoevaporation (citations from Nakatani). 
The overall effect of internal photoevaporation on disk dispersal has remained largely unconstrained, as investigations continue to debate the overall, relative importance of each stellar flux energy regime (e.g. far-ultraviolet, extreme-ultraviolet, and X-ray) on resultant mass-loss rates and disk lifetimes.
%In general, these efforts are also difficult to compare directly given adoptions of different stellar models, varying assumptions of grain sizes and distributions, and disk chemistries. 
Recent work has argued the impact of EUV radiation is limited, given its absorption over small column densities that does not allow for penetration into the high-density disk midplane (\citealt{2010MNRAS.401.1415O}; \citealt{2019MNRAS.487..691P}). In general, the wind mass-loss predictions from EUV-dominated photoevaporation studies (\citealt{1994ApJ...428..654H}; \citealt{2013ApJ...773..155T}) have also been lower by an order of magnitude or more than that from FUV and X-ray investigations (\citealt{2009ApJ...699.1639E}; \citealt{2009ApJ...690.1539G}; \citealt{2010MNRAS.401.1415O}, \citeyear{2012MNRAS.422.1880O}; \citealt{2015ApJ...804...29G}). Given this and the nature of our observations, we choose to mainly interpret our results in the context of current predictions from X-ray driven disk photoevaporation modeling.

Wind mass-loss rates from X-ray driven photoevaporation have been studied in growing detail, with a library of models covering the observed parameter space of both stellar X-ray properties as well as disk metallicity (\citealt{2019MNRAS.487..691P}; \citealt{2019MNRAS.490.5596W}). \citealt{2009ApJ...699.1639E} found that the stellar luminosity in the soft (0.1-1 keV) X-ray band drives the bulk of the photoevaporative wind, as harder X-rays are unable to provide enough disk heating to unbind their gaseous components despite deeper penetration (\citealt{2017ApJ...847...11W}; \citealt{2018ApJ...865...75N}). \citealt{2021MNRAS.508.1675E} derive a functional form (Equation~\ref{mdotequation}) for the total expected mass loss from X-ray photoevaporation (given a soft X-ray luminosity down to log L$_{x}\sim$28.5) and find their estimates are in agreement with previous modeling (\citealt{2012MNRAS.422.1880O}; \citealt{2019MNRAS.487..691P}). Prefactor constants have values of a$_{S}$ = -1.947 x 10$^{17}$, b$_{S}$ = -1.572 x 10$^{-4}$, c$_{S}$ = -2.866 x 10$^{-1}$, and d$_{S}$ = -6.694, respectively.    

%Subsequent wind mass loss estimates from soft X-rays (\citealt{2021MNRAS.508.1675E}) notably span almost two orders of magnitude at the low luminosities observed for the Peter Pan sample (L$_{x}\sim$10$^{28}$-10$^{29}$). 

\begin{equation}\label{mdotequation}
\text{log} \dot{M}_W (L_{X,\text{soft}}) = a_S~\text{exp} \left(\frac{(\text{ln}(\text{log} L_{X,\text{soft}}) - b_S)^{2}}{c_S} \right) + d_S
\end{equation}

\begin{figure}
    \includegraphics[width=\linewidth]{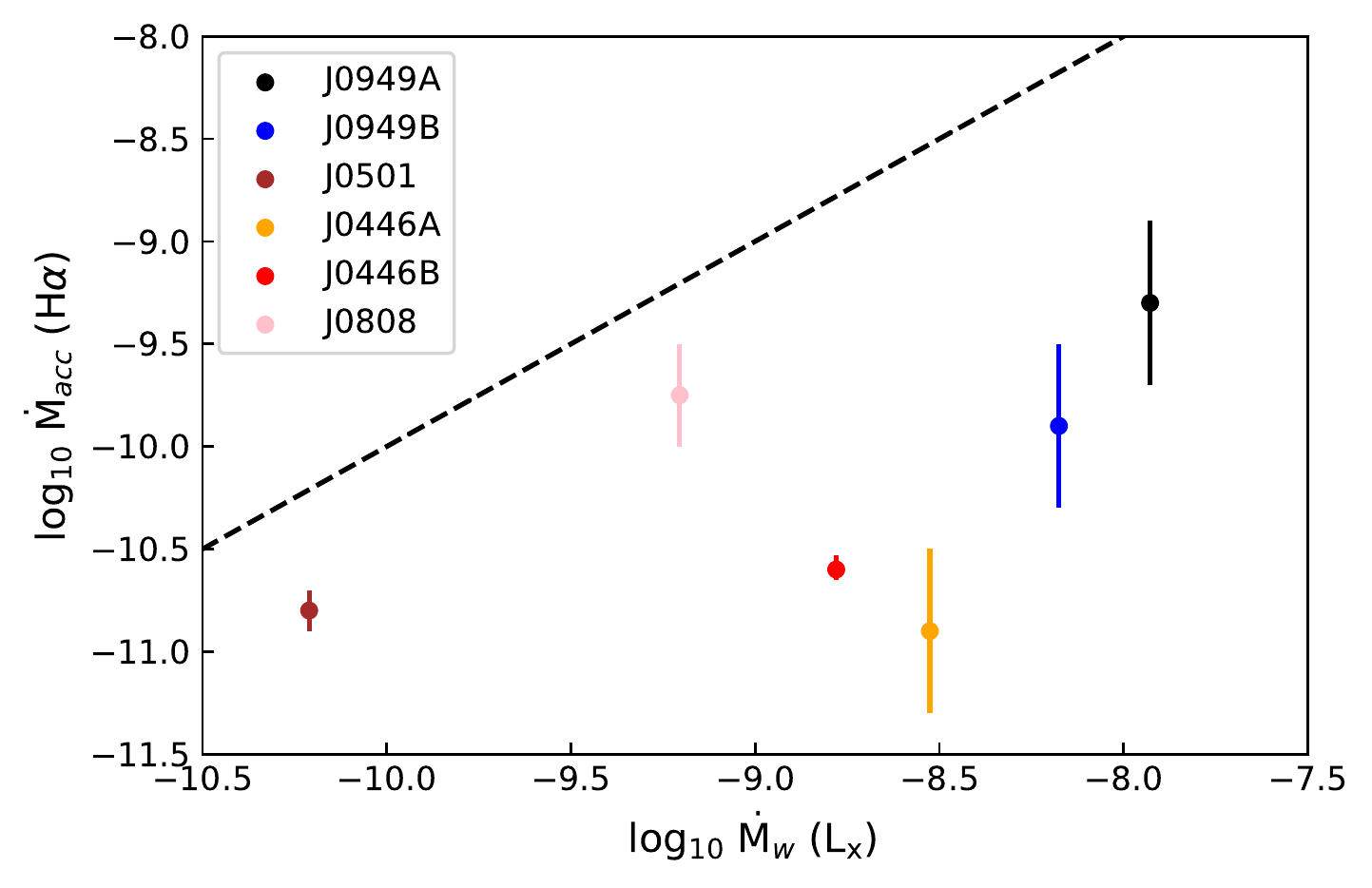}
    \caption{Comparison of the expected mass loss rates between accretion and X-ray driven photoevaporative disk winds for our Peter Pan sample. Values for the former are reported in \citealt{2020ApJ...890..106S} while values for the latter are computed from the functional form (Equation~\ref{mdotequation}) derived in \citealt{2021MNRAS.508.1675E}. We indicate the 1-1 relation with a black dashed line.}
    \label{fig:massrate_comp}
\end{figure}

Using our best fit models in the 0.8-3.0 keV band, we extrapolate to determine the 0.1-1.0 keV X-ray luminosities of our Peter Pan sample and calculate their wind mass-loss rates from Equation~\ref{mdotequation} (as detailed in \citealt{2021MNRAS.508.1675E}). We report these values in Table~\ref{tab:mdotestimates}, conservatively estimating an uncertainty of $\sim$1 dex. We find our observed X-ray luminosities predict appreciable wind mass-loss rates that also span a large range (over 2 dex) across our entire sample. We note these values are effectively lower limits given their X-ray luminosities have decreased throughout their pre-main-sequence evolution. 

These derived wind loss values track the flow of inner disk material ejected into the outer regions of the disk, in contrast to the on-going stellar accretion, which funnels inner disk material to the stellar photosphere. As the mass accretion rate decreases and becomes comparable to the wind mass loss rate, the opening of a gap in the inner disk is expected, clearing the inner disk of material on rapid timescales ($\lesssim$1 Myr, \citealt{2011ARA&A..49...67W}). We compare our X-ray photoevaporative mass loss rates to the Peter Pan accretion rates (derived in \citealt{2020ApJ...890..106S} from H-alpha measurements) to gauge this critical moment of inner disk evolution in Figure~\ref{fig:massrate_comp}. Overall, we find X-ray photoevaporative mass loss rates that are larger than those derived from accretion. Their relative difference could be a consequence of numerous interpretations, which we discuss below.

The relatively low mass loss rates derived from accretion are not surprising given the Peter Pans represent a sample of fairly evolved low-mass stars. \citealt{2022MNRAS.509...44W} have shown that lower disk viscosities, which slow the overall accretion, can allow for longer primordial disk lifetimes (upwards of $\sim$50 Myr for estimated disk properties around stars of $\lesssim$0.5M\textsubscript{\(\odot\)}). However, assuming the rates derived in Equation~\ref{mdotequation} are correct, these results suggest gaps are expected for all of our Peter Pan systems. The two-component SED fits found for J0808 (\citealt{2018MNRAS.476.3290M}; \citealt{2019ApJ...872...92F}) could be evidence of a gap in a Peter Pan disk supporting this interpretation. %While this evidence has not been observed for the remaining Peter Pan objects in our sample, we speculate this could possibly be the case for J0501, the other system for which we see comparable mass loss rates between accretion and X-ray photoevaporation. 

The stronger outliers observed in J0446A/B and J09494A/B are more difficult to explain. We note the greater relative difference in the mass loss rates for J0446A and J0446B could be evidence of H-alpha emission sourced from stellar activity and not accretion. These systems exhibited H-alpha equivalent widths (10-17\AA) and velocity widths (210-240 km/s) that bordered the criteria for an accretion interpretation (\citealt{2003ApJ...592..282J}; \citealt{2009A&A...504..461F}). While our X-ray results may argue against the Peter Pan status of these systems, this explanation is unable to resolve the tension with J0949A, which is much more clearly accreting given its strong (110\AA) and broad ($\sim$370 km/s) emission \citep{2020ApJ...890..106S}. 

X-ray photoevaporative models predict this disk gap phase occurs at an age of $\sim$10-20 Myr (\citealt{2021MNRAS.508.1675E}; \citealt{2021MNRAS.508.3611P}) and constitutes the last $\sim$25\% of the disk lifetime (for a stellar X-ray luminosity of 10$^{29}$ erg s$^{-1}$), leading to a quick dispersal of the remaining disk gas in only a few Myr (\citealt{2021MNRAS.508.1675E}). These short remaining primordial disk lifetimes would be in tension overall with the suspected Peter Pan ages ($\sim$40 Myr). Given it is also unlikely to have caught these systems at a point at which X-ray photoevaporative rates have spiked, we argue the $\dot{M}_W$--L$_{x}$ relation found by \citealt{2021MNRAS.508.1675E} is likely inaccurate for our Peter Pan sample. The discrepancy may be explained by the fact that our observed Peter Pan luminosity range (log $L_{x}$ $\sim$ 27.7--29.1) extends much lower than the range tested in the models of \citealt{2021MNRAS.508.1675E} (log $L_{x}$ $\sim$ 29--31). 

Alternatively, our derived X-ray photoevaporative wind mass loss rates could be overestimated in the case of mechanisms that shield the inner disk from the stellar X-ray radiation. It has been hypothesized that hard X-ray photons could be screened from interacting with the disk via a molecular magneto-hydrodynamic inner disk wind, driven by the accretion-powered active stellar magnetic field \citep{2020ApJ...903...78P}. This mechanism, however, is mainly predicted for strong classical accretors (10$^{-8}$ Msun yr$^{-1}$) and is likely inconsistent with early Peter Pan evolution, given their relatively low masses.

Lastly, we note recent findings have suggested younger ages for the Carina and Columba moving groups, ranging from 13 \citep{2021MNRAS.500.5552B} to 22 \citep{2019AJ....157..234S} Myr. These ages could reduce the tension in their current classification, given a disk e-folding timescale of $\sim$4-5 Myr (as observed for K-type stars, e.g. \citealt{2016MNRAS.461..794P}) could explain a small fraction of $\sim$20 Myr primordial disks. However, these results are in disagreement with the lithium depletion boundary analysis of \citealt{2018MNRAS.476.3290M}, which indicate a suspected age of $\sim$40 Myr for J0808. These younger moving group ages would also not be able to explain the Peter Pan disk candidates identified in other moving groups, such as 2MASS J02265658–5327032 in the $\sim$45 Myr Tuc-Hor \citep{2020ApJ...890..106S} moving group and 2MASS J15460752-6258042 in the $\sim$55 Myr Argus moving group \citep{2020MNRAS.494...62L}. Continued efforts to determine more exact ages for these moving groups are needed.

Overall, our findings question the impact of X-ray photoevaporative disk winds as the primary source of disk dispersal in dM stars. We note photoevaporative winds driven by the host stellar FUV radiation cannot be ruled out from our observations. It has been suspected a large fraction of FUV photons are sourced from the accretion shocks, with a broad range of FUV luminosities ($10^{-6}$L\textsubscript{\(\odot\)} $\lesssim$ L$_{FUV}$ $\lesssim$ L\textsubscript{\(\odot\)}) observed in active classical TTS (\citealt{1998ApJ...492..323G}; \citealt{2012ApJ...744..121Y}). Therefore, Peter Pan disks may be a consequence of the low FUV flux incident on the disk of low-mass dM stars given their relatively low levels of accretion over the course of their pre-main-sequence evolution. Future FUV observations of low-mass dM stars will be critical to confirm this behavior. If FUV photoevaporation is indeed critical for dM stars, however, it remains unclear why more Peter Pan disks have yet to be discovered. %Therefore, these results may serve as evidence for the importance of host stellar FUV radiation in low mass  low mass regime of the  %However, if FUV ultimately dominates disk dispersal for dM stars, but it remains unclear exactly how these primoridal disks have survived to their advanced ages despite prolonged FUV exposure. 

% lifetime estimate ideas?

% From these rates, we also compute expected disk lifetimes assuming a disk gas mass of 0.1Msun (derived using the prescription in \citealt{2022MNRAS.509...44W} assuming a Peter Pan stellar mass of 0.3Msun). 

%= 10^{a + (b * \text{log}M\textsubscript{\(\odot\)})} \left (\frac{t}{1 \text{Myr}} \right)^{-2/5}

\begin{table*}
\centering
\caption{Mass Loss Rate Estimations}
\label{tab:mdotestimates}
\begin{tabular}{ccccc}
\hline \hline
Source & L$_{x}$ (0.1-1 keV) & log (M$_{acc}$)\tablenotemark{a} & log (M$_{w}$)\tablenotemark{b} \\
 & erg/s & Msun / yr & Msun / yr \\
\hline
J0949A & 9.77E+28 & -9.3 & -7.8 \\
J0949B & 2.81E+28 & -9.9 & -8.2 \\
J0501 & 4.15E+27 & -10.8 & -10.2 \\
J0446A & 2.12E+28 & -10.9 & -8.6 \\
J0446B & 1.88E+28 & -10.6 & -8.8 \\
J0808 & 1.17E+28 & -9.75 & -9.2 \\
\hline
\end{tabular}
\tablenotetext{a}{Estimated in \citealt{2020ApJ...890..106S} from H-alpha observations.}
\tablenotetext{b}{Computed using the $\dot{M}_W$--L$_{x}$ relation derived by \citealt{2021MNRAS.508.1675E} (Equation~\ref{mdotequation}).}
\end{table*}

% In the case of J0808 and J0501, our derived X-ray photoevaporation mass loss rates are roughly comparable in magnitude to the mass loss rates probed by H-alpha given our conservative estimate for the uncertainty in our photoevaporative mass loss rates ($\sim$1 dex). These systems have likely developed an inner disk gap and are now in the final phases of their disk dispersal. For the remainder of our sample, we find large X-ray photoevaporative mass loss rates relative to the H-alpha mass loss estimates. These sources  that imply rapid gaseous disk dispersal.    

% These efforts have now begun exploring the lowest mass regimes, studying the Peter Pan like-masses of %$\eq$0.3M\textsubscript{\(\odot\)}. 

% Our results in Sections~\ref{subsec_mstar_results}-\ref{subsec_modelcomp_results} find X-ray luminosities 

% Relatively low levels of X-ray activity were posited as one possible explanation for the suspected long (45 Myr) primordial disk survival times in Peter Pan systems (cite Silverberg?). However, our results in Sections~\ref{subsec_mstar_results}-\ref{subsec_modelcomp_results} indicate X-ray luminosities that are not only consistent with other evolved, mid type M stars but also in some cases larger than those considered in recent modeling (cite WZ). 

%and only recent investigations have begun incorporating the additional effects from strong magnetic fields expected for young, active pre-MS stars. 

\subsection{The Impact of External Photoevaporation Mechanisms on Peter Pan Disk Longevity}\label{subsec_external_disc}

The intense FUV and EUV fields radiating from the rare, massive O and B stars in a dense star forming region serve as another source of disk dispersal by driving a photoevaporative wind in the outer circumstellar disk. The overall impact and efficiency of this external mechanism on primordial disk longevity is still currently debated. Previous observational studies of young (1-5 Myr) star forming regions have found evidence of a correlation between the projected distance to the most luminous ionizing stellar source and the suspected mass of nearby disks in some cases (e.g. Orion Nebula Cluster: \citealt{2014ApJ...784...82M}, \citealt{2018ApJ...860...77E}; $\sigma$ Orionis: \citealt{2017AJ....153..240A}, \citealt{2019A&A...628A..85V}) but not others (e.g. NGC 2024: \citealt{2015ApJ...802...77M}). Recent N-body simulations have not found any evidence of a correlation (\citealt{2019MNRAS.485.4893N}) even after accounting for projection effects (\citealt{2021ApJ...913...95P}). Similarly, disk lifetimes were not found to be a function of projected distance to the cluster core in IC 1396 \citep{2021AJ....162..279S}. Even so, these investigations do find that the extreme case of disks in close ($\sim$0.5 pc) proximity to massive stars are effectively dissipated, with few disks surviving longer than a few Myr on average. 

The evidence of $\sim$45 Myr old primordial disk systems in our Peter Pan sample inherently implies significantly low levels of photoevaporation from external radiation for prolonged durations. The investigations of \citealt{2020MNRAS.496L.111C} find primordial disks around low mass stars could survive up to $\sim$50 Myr in the presence of low external radiation fields ($<$10 Habing fields (1.6 x 10$^{-2}$ erg cm$^{-2}$ s$^{-1}$), corresponding to an external photoevaporative mass loss rate $\leq$10$^{-9}$ M\textsubscript{\(\odot\)}/yr). A direct quantitative estimation of the local radiation experienced by our Peter Pan systems is not feasible. We note qualitative assessments of the external radiation field, based off of the current 3d positions of the Peter Pan sample relative to other moving group members, are also difficult to interpret. The spatial stellar distribution is not informative of the respective moving group dynamics soon after formation, times at which the stellar density was higher and close encounters more probable. 

Although we are unable to probe these early times ($\lesssim$2 Myr), we note the spatial and kinematic properties of Carina, Columba, and Tuc-Hor are somewhat suggestive of unique dynamical evolution, given their fairly large spatial extents ($\sim$10-15 pc) and particularly low velocity dispersions ($\lesssim$1 km/s) relative to other known moving groups \citep{2018ApJ...856...23G}. Recent evidence has also suggested that nearby moving groups of stars may be spatially much more extended than previously thought. The high-precision stellar kinematics offered by \textit{Gaia} DR2 have revealed numerous co-moving stellar streams throughout the galaxy at distances ranging 80-1000 pc (\citealt{2019AJ....158..122K}). Some of these ``Theia" strings have been identified as relatively young ($\lesssim$100 Myr) and are suspected to be primordial, reflecting the shape of the molecular cloud from which they have formed. In some cases, these streams appear to be interconnected with nearby moving groups, exhibiting similar estimated isochoronal ages and \textit{Gaia} DR2 UVW space velocities (\citealt{2021ApJ...915L..29G}). We test for Theia string membership among our Peter Pan sample and note a consistent match with Theia 113 for J0808. %However, we note the dynamical evolution of moving groups and respective Theia strings is still not well understood.

Although the overall dynamical evolution of moving groups and respective Theia strings is still not well understood, $\sim$2-3 crossing times are still predicted for the stars in these moving groups. The expectation is then that all moving group members are equally likely to have experienced nearby proximity to massive members, in contrast with an interpretation of strict isolation resulting in low external photoevaporation. The lack of confirmed massive members as assessed by the BANYAN algorithm (\citealt{2018ApJ...862..138G}) in the moving groups of Carina, Columba, and Tuc-Hor (1 B9 star and no O stars) also argue against the overall impact of external photoevaporation.

If external photoevaporation is most important for disk dispersal in dM stars, we would expect the dM stars nearest the Peter Pan sources to experience similar levels of incident external radiation, and thus also be candidates for harboring long-lived primordial disks. We refer to our list of mid-M stars with high-probability moving group membership (J. Gagné, private communication) to explore this sub-sample. We find the majority of these candidates did not pass the original criteria of the Disk Detective search \citep{2016ApJ...830...84K} as their disk harboring status could not be determined due to low-SNR WISE observations. We cross-reference this sub-sample with \textit{Gaia} DR 3 and report the 5 nearest mid-M stars to the canonical Peter Pan target, J0808, in Table~\ref{tab:MsnearJ0808}. Follow-up near-IR observations of these low-mass stars would help test the extent to which the long-lived disks in our Peter Pan sample could be explained by unusual local, external radiation fields.

\begin{table*}
\centering
\caption{Nearest Carina Mid-Ms to J0808}
\label{tab:MsnearJ0808}
\begin{tabular}{ccccc}
\hline \hline
Source & RA & DEC & SpT & 3D Separation\tablenotemark{a} \\
 & deg & deg &  & pc \\
\hline
2MASS J08111195-6656400 & 122.79962328091591 & -66.94434486171754 & M4.0 & 8.3 \\
2MASS J08441995-6158424 & 131.08297932821128 & -61.97836028331041 & M4.9 & 13.2 \\
2MASS J08122535-6852061 & 123.105464898878 & -68.8682618587011 & M7.0 & 15.5 \\
2MASS J07151705-6555486 & 108.82103844623693 & -65.93003833077316 & M4.3 & 17.0 \\
2MASS J07550342-6717478 & 118.76411975737 & -67.2964623058796 & M4.7 & 19.5 \\
\hline
\end{tabular}
\tablenotetext{a}{Estimated from \textit{Gaia} DR 3 positions and parallaxes.}
\end{table*}

% Such low external radiation fields are predicted to be particularly rare ($<$.001\%) for young embedded stellar clusters (\citealt{2008ApJ...675.1361F}). However, this may not be the case for the the moving groups the Peter Pan systems reside in, which harbor a much lower average stellar density.

Overall, we do not find evidence suggestive of lower external photoevaporation experienced by our Peter Pan sample relative to other moving group members. However, we acknowledge that many of the aspects of both early moving group dynamic and the efficiency of external photoevaporation in the regime of low stellar mass, remain unconstrained. If Peter Pan disk longevity is a consequence of an usually low external FUV radiation field, we predict Peter Pan systems would likely require unusually high amounts of self-shielding during their early evolution to survive to their current suspected ages of 45 Myr and explain their overall rarity. 

%it remains unclear the extent to which our Peter Pan sample have experienced prolonged, unusual levels of isolation relative to the rest of the M star population in their respective moving groups. Our qualitative assessment suggest relative isolation that likely amounts to modest levels of external photoevaporation currently being experienced by our Peter Pan sample. Given the importance of the (unconstrained) surrounding stellar density and clustering of these moving groups at early times, we predict Peter Pan systems would likely also require relatively higher amounts of self-shielding during their early evolution to survive to their current suspected ages of 45 Myr. 
\section{Summary and Conclusion} \label{sec:summary_conclusion}

We present new X-ray observations for our sample of 6 recently identified Peter Pan stars, systems with strong evidence of harboring primordial disks at the advanced ages of $\sim$45 Myr. Our results in Section~\ref{sec:results} reveal the X-ray characteristics of these sources for the first time. We summarize our main results below. 

\begin{itemize}
    \item We observe Peter Pan X-ray emission similar to that observed in young, weak-lined TTS (Figure~\ref{fig:TTS_look}). Their derived X-ray luminosities have an overall large dispersion, spanning roughly 1.5 dex (log L$_{x}$ = 27.7-29.1). We do not find evidence of their X-ray luminosities correlating with their H-alpha equivalent widths, arguing against accretion-driven X-ray suppression being responsible for their extended disk lifetimes. 
    \item Our derived Peter Pan X-ray luminosities are consistent with that measured for field dM stars of similar spectral type and age (Figure~\ref{fig:Mstarcomp}). We conclude Peter Pan disk lifetimes are likely not a consequence of central stars with lower L$_{x}$/L$_{bol}$. 
    \item In some cases, our observed X-ray luminosities differ from those used in the recent modeling of \citealt{2022MNRAS.509...44W} by up to 0.5 dex (Figure 3). It remains unclear the extent to which Peter Pan x-ray luminosities are being over or under predicted, given our large measurement uncertainties, overall small sample size, and the large intrinsic scatter in observed x-ray luminosities.  
    \item Our derived X-ray photoevaporative mass loss rates (Section~\ref{subsec_internal_disc}) are in most cases much larger than that estimated by previous H-alpha measurements. These measurements may suggest our Peter Pan systems are in the final phase of their overall disk lifetime, predicted to occur at $\sim$10-20 Myr (\citealt{2021MNRAS.508.1675E}, \citealt{2021MNRAS.508.3611P}). Given the disparity between this timescale and the suspected ages of Peter Pan systems ($\sim$45 Myr), we argue the wind mass loss relation of \citealt{2021MNRAS.508.1675E} is inaccurate for our Peter Pan sample. Overall, these findings call into question the impact of X-ray photoevaporation on disk dispersal for dM stars and may suggest the importance of FUV-driven disk dispersal mechanisms. Given the size of our sample, additional FUV and X-ray observations of low-mass dM stars are needed to help confirm this result.
    \item Our qualitative assessment of the surrounding Peter Pan environments (Section~\ref{subsec_external_disc}) does not predict unusually low levels of external photoevaporation relative to other respective moving group members. However, given this assessment does not inform of early moving group dynamics, the overall impact of external photoevaporation remains unclear. If Peter Pan disk longevity is a consequence of an usually low external FUV radiation field, we predict that Peter Pan systems would likely have to have experienced unusual amounts of self-shielding to explain their current rarity. 
\end{itemize}

\begin{acknowledgments}

This research has made use of the SIMBAD database and the VizieR catalogue access tool, operated at CDS, Strasbourg, France. This research has also made use of NASA's Astrophysics Data System and ds9, a tool for data visualization supported by the Chandra X-ray Science Center (CXC) and the High Energy Astrophysics Science Archive Center (HEASARC) with support from the JWST Mission office at the Space Telescope Science Institute for 3D visualization. Multiple Python libraries aided the analysis of our data including matplotlib, a Python library for publication quality graphics \citep{Hunter:2007}, SciPy \citep{Virtanen_2020}, NumPy \citep{harris2020array}, and Astropy, a community-developed core Python package for Astronomy \citep{2013A&A...558A..33A, 2018AJ....156..123A}. IRAF is distributed by the National Optical Astronomy Observatory, which is operated by the Association of Universities for Research in Astronomy (AURA) under cooperative agreement with the National Science Foundation \citep{1993ASPC...52..173T}. These acknowledgements were compiled using the Astronomy Acknowledgement Generator.

The scientific results reported in this article are based on observations made by the Chandra X-ray Observatory. This research has made use of software provided by the Chandra X-ray Center (CXC) in the application packages CIAO \citep{2006SPIE.6270E..1VF} and Sherpa \citep{2007ASPC..376..543D}.

\facility{CXO (ACIS)}
%\software{Python}

\end{acknowledgments}

\bibliography{Peter_Pan}{}
\bibliographystyle{aasjournal}

\appendix

\section{Best-Fit Spectral Analysis Models}\label{sec:spec_fitting}

\begin{figure}[!ht]
    \includegraphics[width=\linewidth]{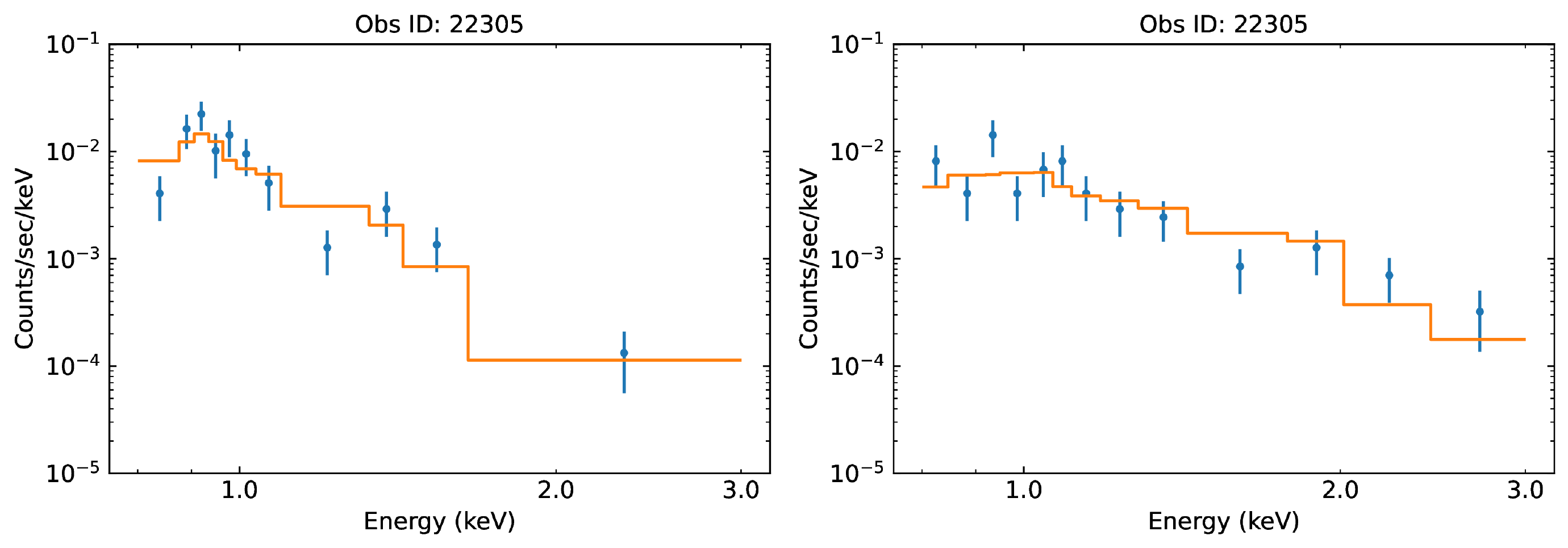}
    \caption{\textit{Chandra}/ACIS X-ray spectra of J0949A (left) and J0949B (right) observed on 2019-12-13. Blue points represent our binned X-ray count data. The orange line represents our best-fit spectral model (see Section~\ref{sec:obs_data}). The data is fit without binning using the Cstat statistic, but is shown here binned to 5 counts/bin for display purposes (error bars are $\sqrt{N}$) to guide the eye.}
    \label{fig:0949specfit}
\end{figure}

\begin{figure}[!ht]
    \includegraphics[]{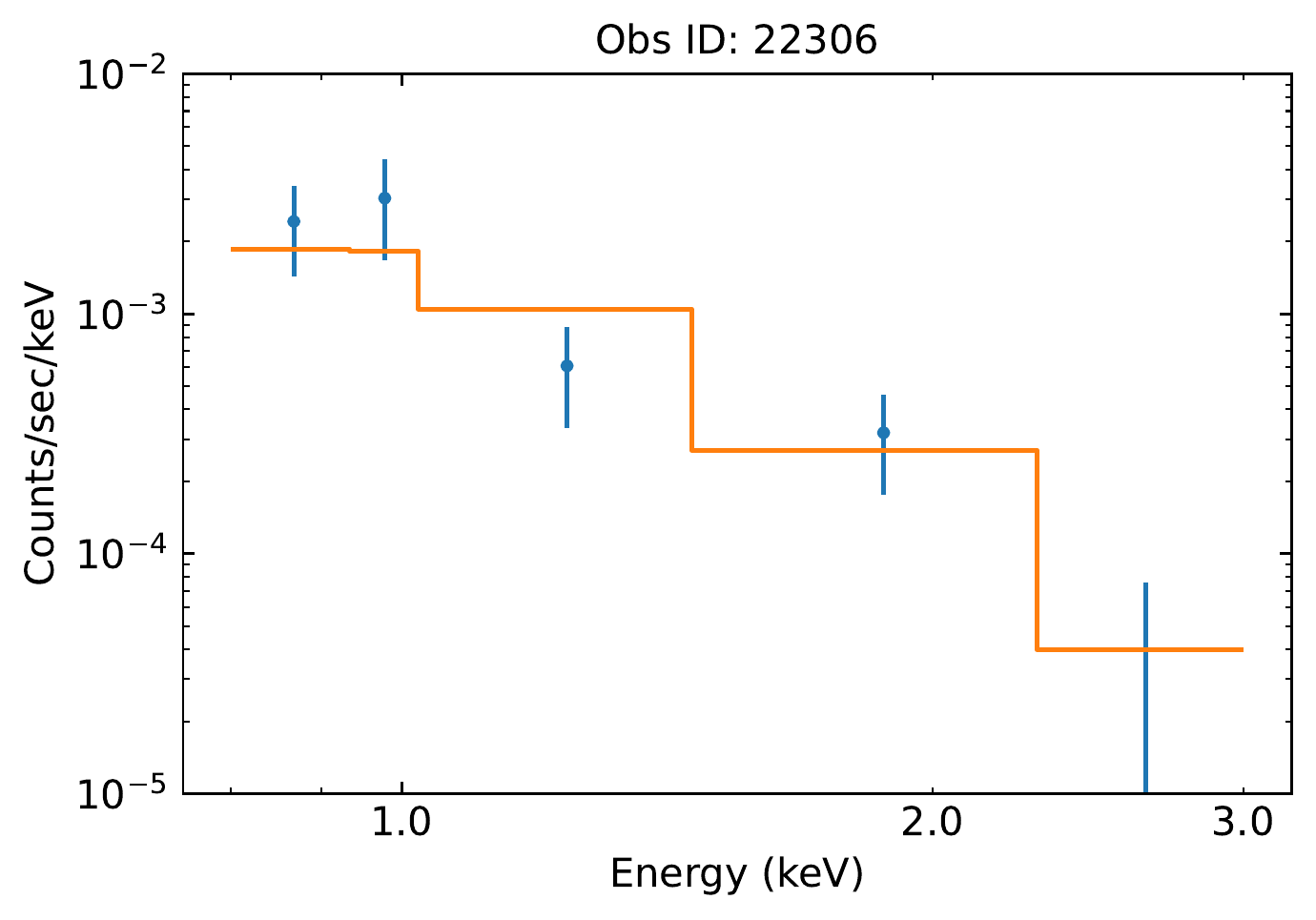}
    \caption{Similar to Figure~\ref{fig:0949specfit}, \textit{Chandra}/ACIS X-ray spectra of J0501 observed on 2019-12-22.}
    \label{fig:0501specfit}
\end{figure}

\begin{figure}[!ht]
    \includegraphics[width=\linewidth]{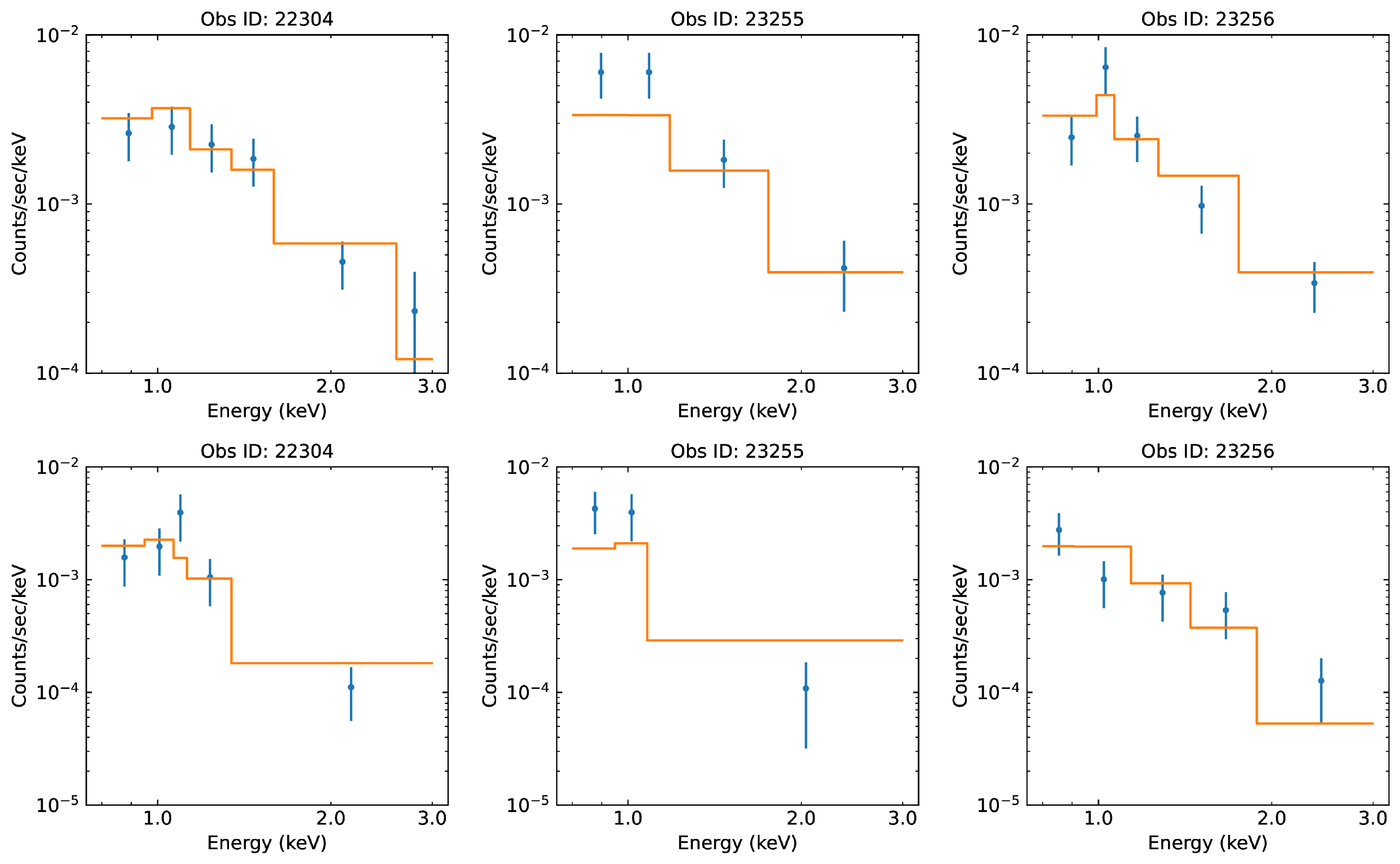}
    \caption{Similar to Figure~\ref{fig:0949specfit}, \textit{Chandra}/ACIS X-ray spectra of J0446A (top row) and J0446B (bottom row) observed on 2020-05-20 (left), 2020-05-21 (middle) and 2020-06-02 (right).}
    \label{fig:0446specfit}
\end{figure}

\begin{figure}[!ht]
    \includegraphics[width=\linewidth]{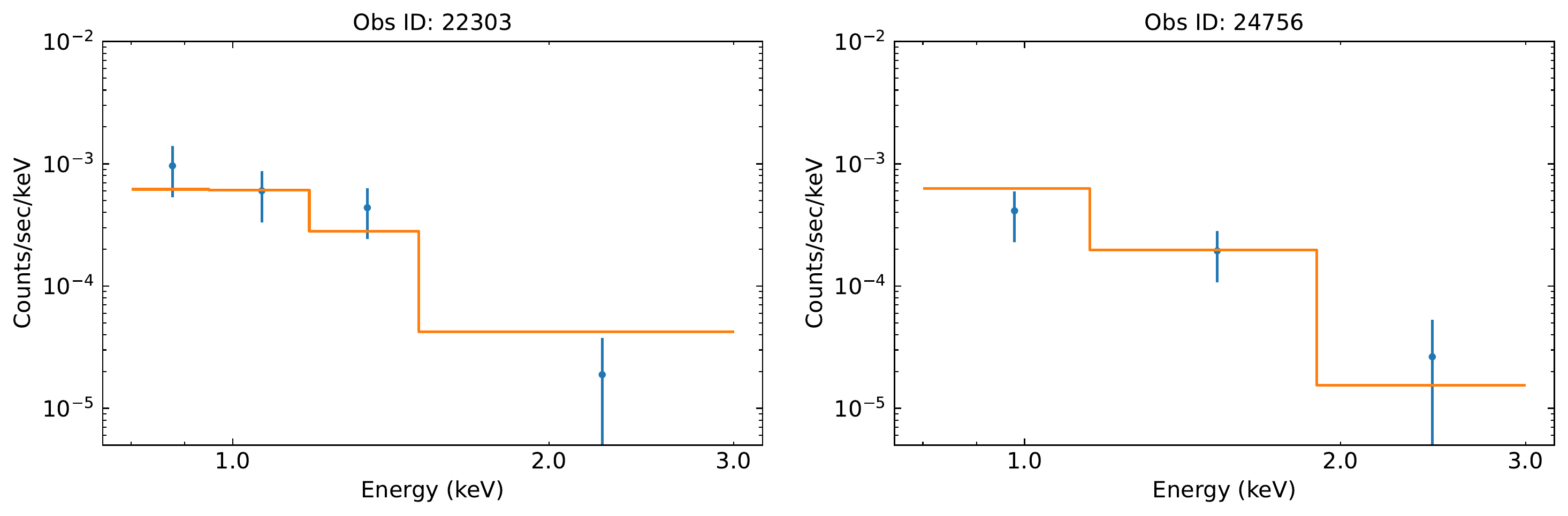}
    \caption{Similar to Figure~\ref{fig:0949specfit}, \textit{Chandra}/ACIS X-ray spectra of J0808 observed on 2021-07-06 (left) and 2021-07-07 (right). }
    \label{fig:0808specfit}
\end{figure}

We describe our spectral fitting procedure in Section~\ref{sec:obs_data}. The resulting best fits from this analysis are shown in Figures~\ref{fig:0949specfit}-\ref{fig:0808specfit}, respectively. We discuss the derived X-ray properties of our Peter Pan sources in Section~\ref{sec:results}. 

\section{M star X-ray Control Sample}\label{sec:xraycomp_sample}

To gauge the extent to which the Peter Pan systems are potentially X-ray under-luminous, we construct a control sample of objects to compare against. For an appropriate comparison, we only consider objects with suspected spectral types in a similar range as that identified for the Peter Pan sources from \citet{2020ApJ...890..106S} (i.e. M4--M7). We also require high ($\gtrsim$95\%) membership probability in the suspected parent moving groups of the identified Peter Pan sources (i.e. Carina, Columba, and Tuc-Hor) as determined by the BAYNAN algorithm \citep{2018ApJ...856...23G}. These populations are expected to be co-eval with derived ages of $\sim$45 Myr \citep{2015MNRAS.454..593B}. 

For this initial sample of 227 objects (J. Gagné, private communication), we simultaneously query the XMM-Newton Serendipitous Source Catalog (4XMM-DR11) as well as the \textit{Chandra} Source Catalog (v.2.0) for XMM/Chandra pipeline-derived X-ray fluxes reported by the HEASARC interface. We consider a coordinate search radius of 5\arcsec~for each object, returning 13 matches. We carefully inspect the X-ray light curves of these matches by eye when available, excluding one source due to an obvious flare feature. To compute X-ray luminosities, we use respective \textit{Gaia} DR3 distances along with the pipeline derived fluxes listed (EP 8 band for XMM-Newton matches, CSC broadband for Chandra matches). We note these energy ranges (0.2-12 keV and 0.5-7.0 keV, respectively) do not exactly line up with our reported 0.3-8 keV luminosities. From our best fit spectral models, we compute our Peter Pan X-ray luminosities in this larger energy band (0.2-12 keV) but find the overall difference is negligible, given the X-ray flux strongly peaks at 1 keV. For this sub-sample of objects, we also compute bolometric luminosities as indicated in Section~\ref{sec:obs_data}. We note we do not consider X-ray fluxes derived by ROSAT observations, given its lower overall sensitivity biases detections towards the brightest, X-ray luminous objects. We report the overall X-ray properties of this control sample in Table~\ref{tab:analog_sample}. 

\begin{table*}
\centering
\caption{Analog M Star X-ray Sample}
\label{tab:analog_sample}
\begin{tabular}{ccccc}
\hline \hline
Source & Moving Group\tablenotemark{a} & Spectral Type & L$_{x}$ (erg/s) & log L$_{x}$/L$_{bol}$ \\
\hline 
2MASS J01275875-6032243 & THA & M4 & 2.3e+28 & -3.33 \\
2MASS J02025502-7259155 & THA & M5.5 & 2.87e+28 & -3.2 \\
2MASS J02153328-5627175 & THA & M4.5 & 9.14e+27 & -3.28 \\
2MASS J03152363-5342539 & THA & M5.2 & 1.9e+28 & -3.27 \\
WOH G 312 & CAR & M4.6 & 1.96e+28 & -3.42 \\
2MASS J08420090-7113216 & CAR & M4.7 & 1.44e+29 & -2.81 \\
2MASS J09421385-5601361 & CAR & M4 & 2.34e+29 & -2.63 \\
UCAC3 101-434911 & THA & M4 & 4.06e+28 & -3.26 \\
2MASS J23143092-5405313 & THA & M5 & 1.12e+28 & -3.3 \\
2MASS J23143092-5405313 & THA & M5 & 4.71e+27 & -3.68 \\
2MASS J05360322-6555191 & THA & M6.5 & 3.27e+27 & -3.4 \\
\hline
\end{tabular}
\tablenotetext{a}{The Tuc-Hor and Carina moving groups are estimated to be 45 Myr old \citep{2015MNRAS.454..593B}.}
\end{table*}

%\nocite{*}

\end{document}